# Geometric heat pumping under continuous modulation in thermal diffusion

Hao-Ran Yan[2], Pei-Chao Cao[1,3,4], Yan-Xiang Wang[1,3,4], Xue-Feng Zhu[2*], & Ying Li[1,3,4*]

[1]*State Key Laboratory of Extreme Photonics and Instrumentation, Key Lab. of Advanced Micro/Nano Electronic Devices & Smart Systems of Zhejiang, Zhejiang University, Hangzhou 310027, China*

[2]*School of Physics and Innovation Institute, Huazhong University of Science and Technology, Wuhan 430074, China*

[3] *International Joint Innovation Center, The Electromagnetics Academy at Zhejiang University, Zhejiang University, Haining 314400, China*

[4]*Shaoxing Institute of Zhejiang University, Zhejiang University, Shaoxing 312000, China*

*e-mail: xfzhu@hust.edu.cn; *e-mail: eleying@zju.edu.cn

**Abstract:** Berry (geometric) phase has attracted a lot of interest and permeated into all aspects of physics including photonics, crystal dynamics, electromagnetism and heat transfer since it was discovered, leading to various unprecedented effects both in classical and quantum systems, such as Hannay angle, quantum Hall effect, orbital magnetism and Thouless pumping. Heat pumping is one of the most prominent and fantastic application of geometric phase in heat transport. Here we derive a general heat pumping theory based on classical diffusion equation and continuous modulation of system parameters in macroscopic thermal diffusion system and obtain a formula which is reminiscent of contact between Berry phase and the Berry curvature. Furthermore, we discuss two cases of non-trivial zero heat flux after one cycle which is fundamentally different from the trivial zero heat flux generated by static zero heat bias in physical nature. Then we analyze the dependence of the effect on the system thermal parameters, including some counterintuitive phenomenon. Finally, under the guidance of this theory, we conduct an experiment to demonstrate the accuracy and effectiveness of our theory and observe the heat pumping effect regardless of the presence and the absence of the thermal bias between two ports of system. In general, our work clearly derives the universal form of heat pumping theory under arbitrary form of the modulation in the macroscopic thermal diffusion system, this is of great significance for better heat energy transport, heat manipulation and so on. It also establishes the foundation of achieving other non-reciprocity devices or topological devices with the aid of spatiotemporal modulation.

The advance of phononics and the rise of the thermal metamaterials inspired the emergence of numerous active areas including thermal cloaking [1-3], thermal camouflage [3-5], thermal non-reciprocity [6], nanoscale thermal conduction [7, 8], non-Hermitian regime [9,10and topological effect [11] in diffusion system. To regulate and manipulate heat and achieve thermal non-reciprocity, heat engines [12,13], thermal diodes [14], thermal logic gate [15], for instance, abundant devices has been developed as electrical counterparts. At the same time, heat can also be applied to coupling to other systems, like thermoelectric effect [9], thermoacoustic oscillation [16], thermo-chemical energy storage [17]. Subsequently, nonlinear materials [18], phase-changing materials [19, 20], spatiotemporal modulation [11,21], and external electromagnetic field [22] are applied to physical systems to demonstrate novel phenomena, such as heat pumping [23], heat shuttling [24], thermal ratcheting [25] and topological phonon Hall effect [26].

After the discovery of the Berry (geometric) phase [27] in 1984, this concept has recently attracted a great deal of interest both in theoretically and experimentally in a variety of field in physics, such as topological physics [27-30] including optics [31,32], acoustic [33-35] and thermal transport [36-38] due to the topological origin of it. Geometric heat pumping is one of applications and promotion of Berry phase in heat transport, it is a device which can generate a current after one cycle based on at least two parameters are slowly and periodically modulated, even though in the absence of the external bias which is the reason why it is considered as one of the most innovative and intriguing phenomena with respect to Geometric phase. The underlying mechanism or applications of the Berry phase in microscopic quantum systems are mature enough with Thouless pumping is the most distinguished [39], for instance, charge pumping which can realize the transport of the charge in the absence of external electromagnetic field and it should be noticed that this transport is quantized owing to the topology of the pump. Another example is the spin pumping as an extension of the charge pumping and it can be described by the Chern number [39] which is topological and quantized as well. Conventional heat transfer is strictly governed by the second law of thermodynamics all the time, that is, heat can only flow spontaneously from higher temperature to lower. Beyond these, research of Berry phase on microcosmic heat transport and thermodynamics in both quantum [40,41] and classical [42] systems are also growing vigorously, while few works concentrate on the macroscopic thermal pumping except a theoretically and experimental exploration of the geometric pumping in macroscopic thermal diffusion system [43], nevertheless, this is under discrete modulation protocol hence there is a certain lack of universality. Far too little attention has been paid to investigating the continuous modulation in diffusion system up to now, so our work fills the crucial gap of this in the macroscopic thermal diffusion system.

In this letter, we derive a general geometric heat pumping theory with several thermal parameters under adiabatic evolution in a two-terminal system under the presence (upper schematic diagram in Figure 1(a)) or absence (lower schematic diagram in Figure 1(a)) of the thermal bias between two reservoirs on either hand of system controlled by classical thermal diffusion equation with the third boundary condition. We pay attention to the case that the instantaneous thermal bias between two reservoirs is strictly zero, after a modulation cycle, there is an anomalous non-zero net heat flux, which is called geometric heat and has completely different physical nature to the dynamical heat flux governed by the laws of the thermodynamics, more importantly, we demonstrate that, due to the presence of the heat pumping effect and a finite temperature difference, the heat transfer after one complete cycle can be enhanced, decreased, and even reversed (lower schematic diagram in Figure 1(a)) with respect to the heat transfer that occurs in the absence of heat pumping effect. Therefore, we can ultimately realize two kinds of non-trivial zero geometric heat by adjusting the thermal bias between two thermal reservoirs and initial phase of the two thermal reservoirs which is modulated in the form of travelling wave. Finally, we proceed an experiment setup to demonstrate our theory and observe the heat pumping effect. To a certain extent, our work successfully promotes the development of the field of thermal non-reciprocity and expands the theoretical ideas for realizing non-reciprocity devices, such as thermal diodes, thermal transistor, other topological thermal devices and so on.

# I. General heat pumping theory under continuous modulation

Consider a one-dimensional diffusion system of length L with two terminals which exchange heat with two thermal reservoirs, respectively. The evolution of temperature field $T(x,t)$ satisfied the classical diffusion equation:

$$\frac{\rho c_p(x) \partial T(x,t)}{\partial t} = \frac{\partial}{\partial x}\left(\frac{\kappa(x) \partial T(x,t)}{\partial x}\right) \quad (1)$$

the boundary condition which follows the Newton's cooling law at two terminals:

$$-\kappa(x)\frac{\partial T(x,t)}{\partial x} = h_0(t)(T_0(t) - T(x,t)), x = 0$$
$$\kappa(x)\frac{\partial T(x,t)}{\partial x} = h_1(t)(T_1(t) - T(x,t)), x = L \quad (2)$$

Where $\rho, C, \kappa(x), h_0(t), h_1(t), T_0(t), T_1(t)$ are mass density, heat capacity, thermal conductivity, heat transfer coefficient, thermal reservoirs of left and right end, respectively. It should be noted that $\Gamma(t) = \{T_0(t), T_1(t), h_0(t), h_1(t)\}$ satisfied $\Gamma(t) = \Gamma(t+\tau)$, then above parameters draw a closed evolutionary path $\Omega$ in one modulation period, the geometric heat which we focus is completely depend on the area enclosed by $\Omega$ and the flux on the area.

We give the general solution of the equation which has the form:

$$T(x,t) = G(x,t) + \sum_{i=1}^{n} c_i(t)\psi_i(x,t)\exp(-W_i(t)) \quad (3)$$

Obviously and importantly, we introduce one another term $G(x,t)$ out of the summation term in this solution, therefore the boundary condition can be resolve into two parts:

$$\kappa(0)\frac{\partial \psi_j}{\partial x}(0,t) = h_0(t)\psi_j(0,t)$$
$$\kappa(L)\frac{\partial \psi_j}{\partial x}(L,t) = -h_1(t)\psi_j(L,t)$$
$$\kappa(0)\frac{\partial G}{\partial x}(0,t) = h_0(t)\big(G(0,t) - T_0(t)\big)$$
$$\kappa(L)\frac{\partial G}{\partial x}(L,t) = -h_1(t)\big(G(L,t) - T_1(t)\big)$$
(4)

In this solution, terms in the summation in Eq (3) also can be solved by applying some appropriate mathematic methods and approximation, see Supplementary Material for more details.

After these, we can introduce the integral representation of the overall heat:

$$Q(x=0) = -\int_0^{\tau_p} \kappa A(t)dt + \int_0^{\tau_p} \left\langle \psi_1(t) \left| \frac{\partial}{\partial t} G(x,t) \right\rangle F(t)dt \quad (5)$$

We divide this result into two parts by their different physical nature:

$$Q_{Ove} = Q_{Dyn} + Q_{Geo}$$

Where $Q_{Dyn}$ is the accumulation of steady current and be everywhere while $Q_{Geo}$ can be considered as a result of the adiabatic evolution of the system parameters or the consistency of time scale of variation in external conditions and system's response, this is also consistent with the tendency of its variation with the period of evolution, that is, at the beginning, it increases over the period, when the period is long enough, it remains a constant and no longer changes with the period, as demonstrated by solid black line and black triangle symbols representing simulation results in Figure 2(d) .

The dynamical term is in the form of a definite integral, which is proportional to the thermal bias between two reservoirs, naturally, if the bias is zero, dynamical heat flow vanishes, there is only geometric flow which can be converted to surface integral with Green's formula:

$$Q_{Geo} = \iint \left(\frac{\partial F(\tau)}{\partial G}\frac{\partial E(\tau)}{\partial h} - \frac{\partial F(\tau)}{\partial h}\frac{\partial E(\tau)}{\partial G}\right) dGdh \tag{6}$$

Where anti-symmetry tensor $S_{Geo} = \frac{\partial F(\tau)}{\partial G}\frac{\partial E(\tau)}{\partial h} - \frac{\partial F(\tau)}{\partial h}\frac{\partial E(\tau)}{\partial G}$ and $Q_{Geo}$ easily remind us of Berry curvature and Berry phase which is independent on evolutionary process and duration, only dependent on evolutionary trajectory of parameters and the flux through a surface enclosed by the trajectory.

As shown in Figure 2(b), The parameters are modulated in the form of traveling waves with period, it is noteworthy that $\phi$, the initial phase of G(t) can't be equal to 0 or an integer multiple of pi to enclose an area of finite size in the parameter space. The magnitude of the geometric curvature $S_{Geo}$ in the 2D parameter space is shown in Figure 2(c), the elliptical solid line represents the modulation trajectory of the parameter in one period, $\phi = \frac{\pi}{2}$ under this circumstance, the flux of $S_{Geo}$ over the elliptical is the geometric heat $Q_{geo}$.

## II. Two cases of non-trivial zero heat

According to the second law of thermodynamics, heat always flows spontaneously from higher temperature to lower. Nevertheless, in our heat pumping theory, we firstly demonstrate that there be a non-zero heat flow after one cycle even between two identical reservoirs by every instant under certain prerequisites mentioned above. Without loss of generality, we set the form of the parameters as trigonometric function, and two thermal reservoirs is not completely identical, more clearly, they differ by a non-zero constant, thereby produce an extra contribution called dynamical heat which is almost proportionally to the constant except for the geometric contribution mentioned above, this extra heat can enhances, reduces, even reverse the direction in the absence of the constant, as shown by the gray line in Figure 3(a), so we can always evaluate a constant of appropriate magnitude so that the dynamical and the geometric contribution cancel out, that's the first kind of non-trivial zero heat. The result of the theory and the simulation is shown in the Figure 3(a), there we use COMSOL Multiphysics® 6.2 for modeling and simulation.

Then divert our attention to the second case of non-trivial zero heat which is achieved by adjusting the initial phase of two identical thermal reservoirs $\phi$. We first analyse this phenomenon theoretically. For a given set of $\{h_0(t), h_1(t)\}$ of traveling wave form with initial phase $\varphi, \varphi + \Delta\varphi$, and the value of $\phi$ ranges from 0 to 2pi. Notice that in our theory, when is no thermal bias between two reservoirs, $G(t)$ contributes to the result only in the inner product $\left\langle \psi_j(t) \left| \frac{\partial}{\partial t} G(t) \right.\right\rangle$, it can be expressed more clearly as $\langle \psi_j(t) | 1 \rangle \frac{\partial}{\partial t} G(t)$, and obviously the $\frac{\partial}{\partial t} G(t)$ is still a traveling wave with the same $\phi$, then we view result $Q_{Geo}(\phi)$ as the function of $\phi$, that is, $Q_{Geo}(\phi)$ which is obviously continuous with respect to $\phi$. Without loss of generality, for $\phi = 0$, $Q_{Geo}(\phi)$ takes a non-zero value $Q_{Geo}(0)$, then we take $\phi = \pi$, $Q_{Geo}(\phi)$ takes another non-zero value $Q_{Geo}(\pi) = -Q_{Geo}(0)$, so according to the continuity of the function with respect to $\phi$, there must be a $\phi_0$ between zero and pi that make $Q_{Geo}(\phi)$ equals zero, and of course $\phi_0 + \pi$ is also satisfied. For each different $\Delta\varphi$, we can all find the only two $\phi_0$ that makes $Q_{Geo}(\phi)$ equals to zero, as shown in Figure 3(b). This non-trivial zero heat is the result of the counteraction of the fluxes of the geometric curvature over the closed area in the parameter space, in a geometric sense.

## III. Dependence on system parameters

In this theory, there are numerous parameters that can influence final result to some extent, the most obvious and direct is that $\Gamma(t) = \{T_0(t), T_1(t), h_0(t), h_1(t)\}$ as a set of orthogonal basis vectors in the parameter space can directly change both the geometric curvature and its flux in the parameter space. Modulation period, thermal conductivity, average and amplitude of thermal reservoirs, for instance, other factors can also influence the result more or less.

Theoretically, from boundary condition, solving the eigenfunction of the Hamiltonian to the integral of the result, there is a $h_{0,1}(t)$ in all of them. As the quantization of the ability of heat transfer between the boundary of system and the external thermal reservoirs, we might intuitively conclude that, the heat $Q_{Geo}$ will increase monotonically as $h_{0,1}(t)$ increases, but the truth is, within certain limits, when the average and amplitude of $h_{0,1}(t)$ increases, the result will always decrease. In theory, but not intuition, $W_i(t)$, which characterizes the decay term of temperature field with time, contributes the most to $Q_{Geo}$, specifically $W_i(t)$ is proportional to the square of the wave number of the eigenstate of Hamiltonian, and the wave number is obtained by solving a transcendental equation that includes both $h_0(t)$ and $h_1(t)$, and notice that $W_i(t)$ is placed in the position of exponent e which is natural logarithm, so the average and amplitude of $h_{0,1}(t)$ directly affect the result to a great extent. From a quantitative point of view, the wave number increases as the average and amplitude of $h_{0,1}(t)$ increase, so $W_i(t)$ also increases. In Eq. (7), all the other terms except $\exp(W_i(\tau) - W_i(t))$ change more slowly than $W_i(t)$ with $h_{0,1}(t)$, so these leads to this counterintuitive phenomenon. By contrast, the effect of the thermal conductivity with the result seems more intuitive due to the different variation tendency of the $W_i(t)$ under the increase of the thermal conductivity and the heat transfer coefficient.

As shown in Figure 2(d), Figure 3 and that mentioned above, when the modulation period of the system is long enough, the results no longer change with the length of the period. According to the adiabatic approximation theory, if a given disturbance acts slowly enough on a physical system, then there is enough time for the system to adapt external change and response accordingly, and the physical system will maintain its instantaneous eigenstate at all times, then when the modulation period of the system exceeds a certain value, the response of the system will not change with the length of the period. In theory, just like the previous idea, we think of $Q_{Geo}$ as a function of the period $\eta$ after some proper transformation, that is, $Q_{Geo}(\eta)$. When $\eta$ is large enough, the derivative of with respect to $\eta$ is very close to zero, as shown in Figure 2(d).

Considering the influence of time varying thermal reservoirs on the result, the thermal reservoirs consists a constant term (average) and a oscillating term (amplitude) in the form of a traveling wave, nonetheless, in our formulas, the constant term disappears after $G(x,t)$ is derivative with respect to time t, so only the oscillating term contribute to the result, and $Q_{Geo}$ is proportional to the amplitude of the oscillating term, this can also understood as: because the form of two thermal reservoirs is exactly identical, the static constant term doesn't make any additional contribution to the system, but the oscillation term does, see Supplementary Material for more details.

# reference

1. Y. Li, X. Shen, Z. Wu, J. Huang, Y. Chen, Y. Ni, and J. Huang, Phys. Rev. Lett. 115, 195503 (2015).

2. Y. Li, K. J. Zhu, Y. G. Peng, W. Li, T. Z. Yang, H. X. Xu, H. Chen, X. F. Zhu, S. H. Fan, and C. W. Qiu, Nat. Mater. 18, 48 (2019).

3. Y. Li, W. Li, T. C. Han, X. Zheng, J. X. Li, B. W. Li, S. H. Fan, and C. W. Qiu, Nat. Rev. Mater. 6, 488 (2021).

4. M. Li, D. Liu, H. Cheng, L. Peng, and M. Zu, Sci. Adv. 6, eaba3494 (2020).

5. Y. Li, X. Bai, T. Z. Yang, H. L. Luo, and C. W. Qiu, Nat. Commun. 9, 273 (2018).

6. J. X. Li, Y. Li, P.-C. Cao, M. H. Qi, X. Zheng, Y.-G. Peng, B. W. Li, X.-F. Zhu, A. Alù, and C.-W. Qiu, Nat. Commun. 13, 167 (2022).

7. Y. Guo and M. Wang, Phys. Rep. 595, 1 (2015).

8. L. Yang, M. P. Gordon, A. K. Menon, A. Bruefac, K. Haas, M. C. Scott, R. S. Prasher, and J. J. Urban, Sci. Adv. 7, eabe6000 (2021).

9. Y. Li, Y.-G. Peng, L. Han, M.-A. Miri, W. Li, M. Xiao, X.-F. Zhu, J. L. Zhao, A. Alù, S. H. Fan, and C.-W. Qiu, Science 364, 170 (2019).

10. Y. K. Liu, P. C. Cao, M. H. Qi, Q. K. L. Huang, F. Gao, Y. G. Peng, Y. Li, and X. F. Zhu, Sci. Bull. 69, 1228 (2024).

11. G. Xu, Y. Yang, X. Zhou, H. Chen, H. Alu, and C.-W. Qiu, Nat. Phys. 18, 450 (2022).

12. K. Brandner and K. Saito, Phys. Rev. Lett. 124, 040602 (2020).

13. B. Bhandari, P. T. Alonso, F. Taddei, F. von Oppen, R. Fazio, and L. Arrachea, Phys. Rev. B 102, 155407 (2020).

14. J. Ordonez-Miranda, Y. Y. Guo, J. J. Alvarado-Gil, S. Volz, and M. Nomura, Phys. Rev. Appl. 16, L041002 (2021).

15. Hamed and S. Ndao, Sci. Rep. 10, 2437 (2020).

16. M. P. Juniper and R. I. Sujith, Annu. Rev. Fluid Mech. 50, 661 (2018).

17. M. Angerer, M. Becker, S. Härzschel, K. Kröper, S. Gleis, A. Vandersicke, and H. Spliethoff, Energy Rep. 4, 507 (2018).

18. G. Dai and J. Huang, Int. J. Heat Mass Transf. 147, 118917 (2020).

19. J. Ordonez-Miranda, R. Anufriev, M. Nomura, and S. Volz, Phys. Rev. B 106, L100102 (2022).

20. X. Y. Shen, Y. Li, C. R. Jiang, and J. P. Huang, Phys. Rev. Lett. 117, 055501 (2016).

21. D. Torrent, O. Poncelet, and J. C. Batsale, Phys. Rev. Lett. 120, 125501 (2018).

22. S. Edalatpour, J. DeSutter, and M. Francoeur, J. Quant. Spectrosc. Radiat. Transf. 178, 14 (2016).

23. Z. Wang, J. Chen, and J. Ren, Phys. Rev. E 106, L032102 (2022).

24. I. Latella, R. Messina, J. M. Rubi, and P. Ben-Abdallah, Phys. Rev. Lett. 121, 023903 (2018).

25. S. Wang, C. Zeng, G. Zhu, H. Wang, and B. Li, Phys. Rev. Res. 5, 043009 (2023).

26. F. Zhuo, H. Li, and A. Manchon, Phys. Rev. B 104, 144422 (2021).

27. M. V. Berry, Proc. R. Soc. Lond. A 392, 45 (1984).

28. C. P. Schmid, L. Weigl, P. Grössing, V. Junk, C. Gorini, S. Schlauderer, S. Ito, M. Meierhofer, N. Hofmann, D. Afanasiev, J. Crewse, K. A. Kokh, O. E. Tereshchenko, J. Güdde, F. Evers, J. Wilhelm, K. Richter, U. Höfer, and R. Huber, Nature (London) 593,

385 (2021).

29. D. Smirnova, D. Leykam, Y. Chong, and Y. Kivshar, Appl. Phys. Rev. 7, 021306 (2020).

30. L. Lu, J. Joannopoulos, and M. Soljačić, Nat. Photon. 8, 821 (2014).

31. C. P. Jisha, S. Nolte, and A. Alberucci, Laser Photon. Rev. 15, 2100003 (2021).

32. V. G. Kravets, F. Schedin, R. Jalil, L. Britnell, R. V. Gorbachev, D. Ansell, B. Thackray, K. S. Novoselov, A. K. Geim, A. V. Kabashin, and A. N. Grigorenko, Nat. Mater. 12, 304 (2013).

33. E. Hasman, V. Kleiner, G. Biener, and A. Niv, Appl. Phys. Lett. 82, 328 (2003).

34. H. Nassar, B. Yousefzadeh, R. Fleury, M. Ruzzene, A. Alù, C. Daraio, A. N. Norris, G. Huang, and M. R. Haberman, Nat. Rev. Mater. 5, 667 (2020).

35. M. Muhammad, and C. W. Lim, Arch. Comput. Methods Eng. 29, 1137 (2022).

36. H. Xue, Y. Yang, and B. Zhang, Nat. Rev. Mater. 7, 974 (2022).

37. M. Qi, D. Wang, P.-C. Cao, X.-F. Zhu, C.-W. Qiu, H. Chen, and Y. Li, Adv. Mater. 34, 2202241 (2022).

38. Z. Wang, L. Wang, J. Chen, C. Wang, and J. Ren, Front. Phys. 17, 1 (2022).

39. L. J. Xu, J. Wang, G. L. Dai, S. Yang, F. B. Yang, G. Wang, and J. P. Huang, Int. J. Heat Mass Transfer 165, 120659 (2021).

40. R. Citro and M. Aidelsburger, Nat. Rev. Phys. 5, 87 (2023).

41. M. Josefsson, A. Svilans, A. M. Burke, E. A. Hoffmann, S. Fahlvik, C. Thelander, M. Leijnse, and H. Linke, Nat. Nanotechnol. 13, 920 (2018).

42. S. Seah, S. Nimmrichter, and V. Scarani, Phys. Rev. Lett. 124, 100603 (2020).

43. J. Ren, S. Liu, and B. Li, Phys. Rev. Lett. 108, 210603 (2012).

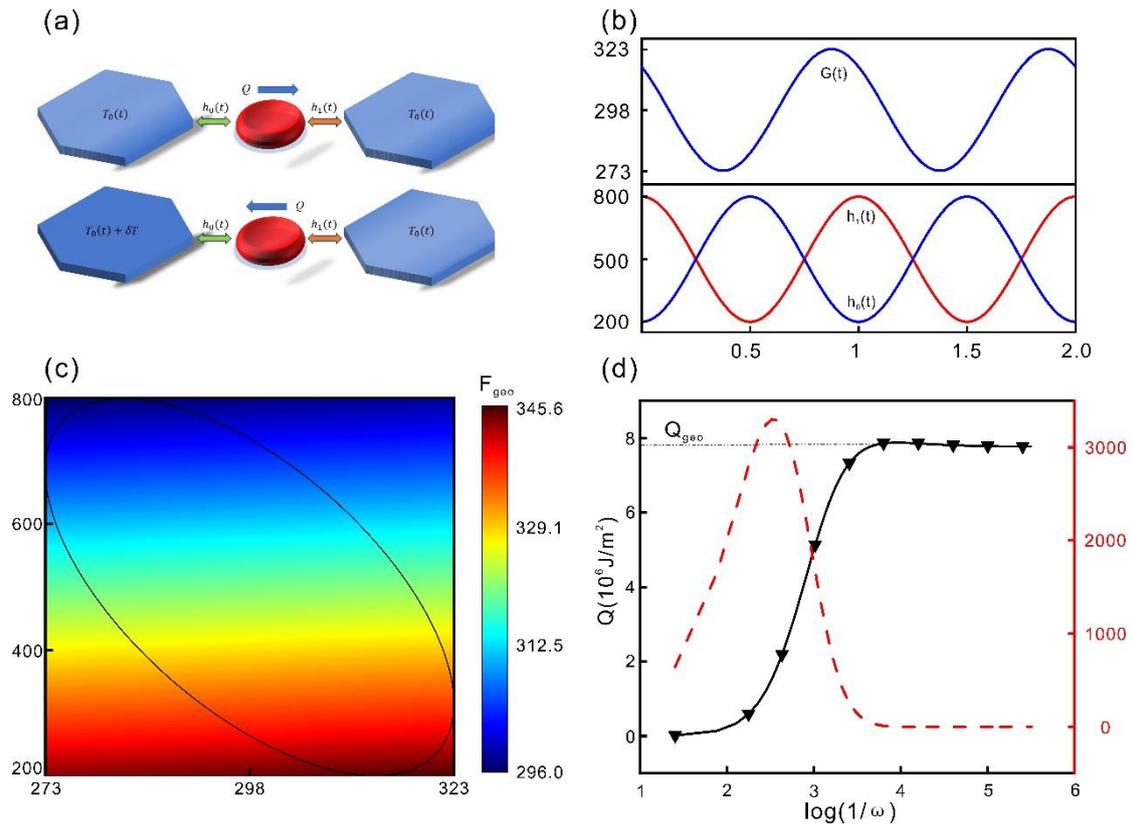

**Figure. 2** Theoretical configuration and result of heat pumping effect. (a) The diagram of two 1D two-terminals system consists of two thermal reservoirs $T_{0,1}(t)$ and central medium under heat pumping effect, thermal reservoirs exchange heat with the medium by following the Newton's cooling law with transfer coefficient $h_{0,1}(t)$. The upper panel represents the non-zero directional geometric heat between two identical reservoirs, the lower panel represents the heat flow from cold to hot. The system parameters in (a) are modulated in the form of traveling wave by (b). (c) The magnitude of the geometric curvature $S_{geo}$ in a 2D parameter space, the solid line of the ellipse is encircled by the parameters during modulation, counterclockwise indicates the positive direction. (d) The dependence of the geometric heat with modulation period, as the period is long enough, the heat gradually approached $Q_{geo}$ and then remain unchanged.

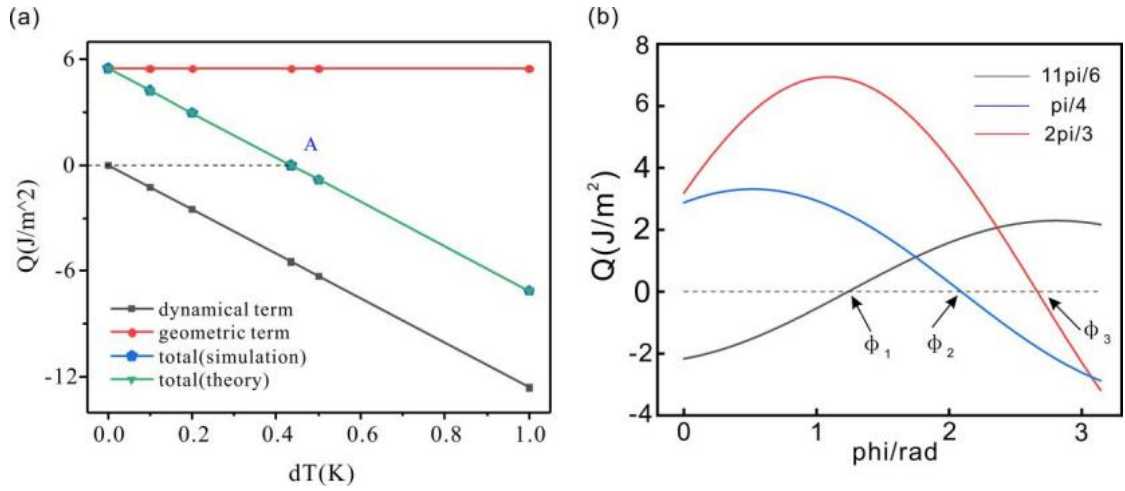

**Figure. 3** Two cases of non-trivial zero heat under heat pumping effect. (a) The point 'A' exhibits the counteraction of the dynamical heat and geometric heat while there is a constant difference between two thermal reservoirs. The error between simulation and theory is less than 0.1%. (b) There is no thermal bias between reservoirs and hence no dynamical heat in the system, another non-trivial zero heat is achieving by changing the initial phase $\phi$ of the thermal reservoirs, obviously we can find two $\phi_0$ that makes $Q_{Geo}$ equals to zero in one complete period, the $\frac{11}{6}\pi$, $\frac{1}{4}\pi$, $\frac{2}{3}\pi$ represent three different cases under various phase difference between $h_0(t)$ and $h_1(t)$.